\documentclass{aa}  

\usepackage{graphicx}
\usepackage{txfonts}
\usepackage{xcolor}
\begin{document}

   \title{Plasma lens with frequency-dependent dispersion measure effects on fast radio bursts}
    \titlerunning{Plasma Lensing Effect on FRB}

   \author{Yu-Bin Wang
          \inst{1}, 
          Xia Zhou
          \inst{2, 3, 4}\fnmsep\thanks{Corresponding author},
          \and
          Abdusattar Kurban
          \inst{2, 4}
          }

   \institute{School of Physics and Electronic Engineering, Sichuan University of Science \& Engineering, Zigong 643000, China \\
   \and
   Xinjiang Astronomical Observatory, Chinese Academy of Sciences, Urumqi 830011, China \\
   \email{zhouxia@xao.ac.cn}
   \thanks{Email of corresponding author}
   \and
   State Key Laboratory of Radio Astronomy and Technology, Beijing, 100101, P. R. China \\
         \and
            Xinjiang Key Laboratory of Radio Astrophysics, 150 Science1-Street, Urumqi 830011, China 
             }

   \date{Received 24 November 2024 / Accepted 11 August 2025}

 
  \abstract{Radio signals propagating through inhomogeneous plasma media deviate from their original paths, producing frequency-dependent magnification effects. In this paper, after reviewing the classical plasma-lensing theory, we have found a fundamental contradiction: the classical model assumes that the distribution of lensing plasma medium is related to the frequency-independent image position; however, our analysis demonstrates that both the image position ($\theta(\nu)$) and dispersion measure (DM$(\nu)$) are inherently frequency-dependent when signals traverse a structured plasma medium.
  We have been able to resolve this paradox by developing a framework that explicitly incorporates frequency-dependent dispersion measures (DMs) following power-law relationships ($\rm DM\propto \nu^\gamma$). Our analysis shows that the signal magnification decreases systematically with decreasing frequency, offering a plausible explanation for the frequency-dependent peak flux densities observed in fast radio bursts (FRBs), particularly in the case of the repeating FRB 180814.J0422+73. Our results suggest these FRBs could originate from the magnetized compact star magnetospheres. By considering these plasma-lensing effects on the sub-pulses of an FRB across different frequencies, we have the ability to more accurately investigate the intrinsic properties of FRBs via precise measurements of radio signals.}
  

   \maketitle
%

\section{Introduction}

Fast radio bursts (FRBs) are enigmatic astronomical phenomena characterized by millisecond-duration radio pulses with remarkably high flux densities in the Jansky range \citep{Lorimer2007}. Over 800 FRBs have been detected to date thanks to a number of radio telescopes worldwide, including some 60 repeating FRBs \citep{Petroff2016,Luo2020,Macquart2020,Amiri2021,Andersen2023,Amiri2024,Sherman2024}. These observations provide valuable opportunities to explore both the nature of FRB source environment and the properties of the intervening medium their signals end up propagating through \citep{Zhang2023}.

Recent observations and theoretical analyses increasingly suggest that many FRBs originate from highly magnetized compact stars \citep{Luo2020,Zhang2023,Mckinven2024}. The polarization angle variations in both some repeating and non-repeating FRBs are similar to pulsar emission \citep{Luo2020,Xu2022,Mckinven2024,Pandhi2024,Amiri2024}, the detection of FRB 20200428 from magnetar SGR 1935+2154 in the Milky Way exhibited characteristics similar to some cosmological FRBs \citep{Andersen2020} and some FRBs, particularly repeating ones, exhibit the "sad trombone" effect, a systematic drift in the arrival times of sub-pulses across different frequencies \citep{Amiri2019,Josephy2019,Day2020,Faber2024}. 

A particularly intriguing feature observed in the sub-pulse components of some FRBs exhibit the rapid decline in peak flux density $P(\nu)$ as their center frequencies decrease \citep{Gajjar2018,Amiri2019,Amiri2020,Luo2020}. For example, FRB 180814.J0422+73 shows decreasing peak flux densities in sub-pulse components at progressively lower center frequencies. This implies that sub-pulse components at lower frequencies have correspondingly lower isotropic-equivalent luminosities ($L_{\rm iso} \propto P(\nu)$; \citealt{Petroff2016})

One proposed explanation for the "sad trombone" effect involves radius-to-frequency mapping in magnetospheric emission models \citep{Wang2019,Lyutikov2020}; however, this interpretation is not universally accepted and alternative explanations exist.
The model assumes the presence of various signals at different frequencies traversing identical propagation paths and same dispersion measure (DM). However, this framework could face challenges in the context of the frequency-dependent luminosity variations. A recent analysis by \citet{Tuntsov2021} proposed an alternative explanation where pulse components initially experience different DM delays before encountering plasma lenses. In this model, standard de-dispersion techniques applied to signals that have traversed frequency-dependent plasma structures \textbf{($\rm DM(\rm \nu)$)} can produce apparent "sad trombone" features without requiring intrinsic emission mechanisms. Enlightened by this model, we speculate that plasma lens with frequency-dependent plasma structures could cause the isotropic equivalent luminosity of some FRBs related with frequency in observations.

Plasma lens, caused by inhomogeneous electron density distributions, can significantly modify radio signal properties, leading to the de-magnification of peak flux density and frequency-dependent path variations \citep{Er2018,Er2022}. These effects depend on multiple factors, including source position, lensing geometry, DM, and signal frequency \citep{Cordes2017,Er2020}. 
Recent observations have revealed evidence for frequency-dependent DM in some sources \citep{Donner2019,Kaur2022}, with limited evidence suggesting these variations might follow approximate power-law relationships in certain cases. Such frequency-dependent DM measurements provide crucially observational supports for plasma-lensing models, as they directly reflect the different electron column densities encountered by signals at different frequencies traversing distinct paths \citep{Donner2019,Tuntsov2021}. Recent studies of FRB signals have shown features that could be consistent with plasma-lensing interpretations \citep{Wang2023,Faber2024}, although other explanations remain viable.

In this paper, we reviewed classical plasma-lensing theory \citep{Clegg1998,Cordes2017,Er2018} and identified a fundamental contradiction in this theory: the lensing plasma medium is assumed to be the distribution related to the frequency-independent image positions; however our analysis results present that both lensing plasma medium and image position are dependent on the frequency of radio signals from a cosmic source. Moreover, the assumption is inconsistent with the frequency-dependent DM observed in repeating FRBs \citep{Tuntsov2021} and pulsars \citep{Donner2019,Kaur2022}. To resolve this paradox, we have developed a comprehensive theoretical framework that explicitly incorporates frequency-dependent paths and the resulting frequency-dependent DM. We demonstrate how this naturally explains some observed properties of FRBs, particularly the systematic frequency dependence of their flux densities. This work focuses specifically on frequency-dependent magnification effects occurring within the duration of individual FRB's bursts. Our analysis suggests that plasma-lensing effects could be used to understand some observational properties of FRB signals.

This paper is organized as follows. Section \ref{sec2} presents the theoretical framework for the plasma lens, reviewing the geometric properties of classical plasma lens and find the contradiction between the assumption and results. Section \ref{sec3} uses the frequency-dependent DM to derive new plasma-lensing equation. Section \ref{sec4} illuminates the frequency-dependent magnification of images and applies our model to observational data, focusing on FRB 180814.J0422+73. The discussion and conclusions are presented in Section \ref{sec5}.

\section{Theoretical framework of the plasma lens}\label{sec2}
\subsection{Basic principles and assumptions}\label{subsec2.1}

We established a Cartesian angular coordinate system to describe the propagation of light rays through a thin plasma-lensing plane. When light rays of frequency, $\nu$, pass through an angular position $\theta$ on this plane, they undergo geometric and dispersive effects compared to the non-lensing case \citep{Clegg1998,Cordes2017,Er2020,Wang2022}. This leads to the two terms of the delay times, namely, the dispersive and geometric delays, respectively \citep{Wagner2020}. The intensity of light ray is also changed due to the magnified or de-magnified effect for plasma lens.

To discuss the physical properties of images conveniently, some approximated principles between the plasma lens and radio signals are employed in this work, as listed below. 
\begin{itemize}
\item Thin-lens approximation: we employed the thin-lens approximation, which can remain valid when the lens transverse scale ($d_{\rm ol}\sigma\sim200\,\rm AU$) is much smaller than the propagation distances ($d_{\rm ol},\,\, d_{\rm os}\sim500\,\rm Mpc$), ensuring that geometric effects dominate over finite thickness corrections. Thus, the incident and exit positions can be assumed at the position with the same transverse distance relative to the lensing center approximately.
\item Source emission characteristics: we considered the source emission characteristics. Some FRBs could originate either from magnetar emission beyond the pulsar light cylinder \citep{Metzger2019,Chen2023} or from coherent radiation bunches in the pulsar magnetosphere \citep{Zhang2023,Kumar2017}. At GHz frequencies, the radiation solid angle of a bunch ($\Omega_{\rm F} > 10^{-4} \, \rm rad^2$) significantly exceeds the solid angle subtended by the plasma lens relative to the source ($\Omega_{\rm PL} = d_{\rm ol}^2\sigma^2/d_{\rm ls}^2 \sim 10^{-6} -10^{-9} \rm \, rad^2$ using parameters defined in later sections). This ensures that bursts at different frequencies arrive approximately parallel at the lens.
\item Stationary and non-rotating lens: we assume the lens within the duration of a FRB is stationary relative to both observer and source, and non-rotating. This ensures the quasi-stationarity of light ray refraction during observation \citep{Wagner2020} and allows us to neglect wave phenomena.
\end{itemize}

These simplifying assumptions make the problem analytically tractable, while preserving the essential physics of frequency-dependent plasma-lensing effects. The following sections build upon this framework to analyze the relationships between image properties and frequency.

\subsection{The formalism of the total time delay}\label{subsec2.2} 

\begin{figure*}
\centering
  \includegraphics[width=\linewidth]{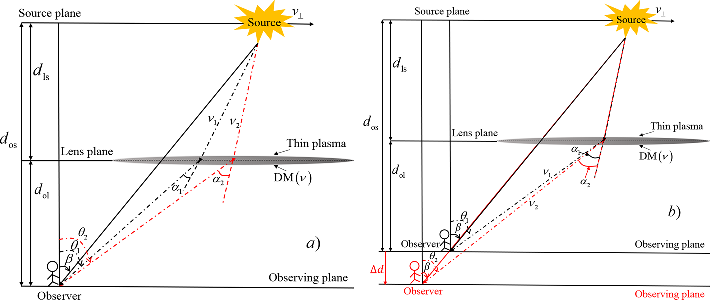}
  \caption{Schematic representation of plasma-lensing geometries. Grey disc stands for plasma lens, and darker region is for higher plasma medium. Panel a: Proposed model, showing frequency-dependent paths through the lens, where signals at frequencies $\nu_1$ and $\nu_2$ (black and red dot-dashed lines) encounter different electron densities. Panel b: Traditional model assuming frequency-independent paths. Solid lines show undeflected paths, $\alpha_i$ and $\theta_i$ denote the deflection angles and image positions ($i =$ 1, 2), while $\Delta d$ indicates the separation between observers.} \label{fig1}
\end{figure*}

The total time delay for radio signals traversing a plasma lens is made up of two components: geometric delay and dispersive delay. The geometric delay ($t_{\rm g}$) arises from the extended path length and is given by \citep{Wagner2020} as
\begin{eqnarray}\label{geometric_delay}
t_{\rm g} =  \frac{(1 + z_{\rm d})}{2 c}\frac{d_{\rm ol}d_{\rm os}}{d_{\rm ls}} \alpha(\theta)^2,
\end{eqnarray}
where $d_{\rm ol}$, $d_{\rm os}$, and $d_{\rm ls}$ represent the observer-lens, observer-source, and lens-source distances respectively, as denoted in Fig. \ref{fig1}, while $\alpha(\theta)$ is the deflection angle at the angular position, $\theta$, representing the deviation of radio signal paths from straight-line propagation due to plasma-lensing effects. The dispersive delay ($t_{\psi}$) results from plasma-induced dispersion and is expressed \citep{Cordes2017,Wagner2020} as
\begin{eqnarray}\label{dispersive_delay}
t_{\psi}  =  \frac{(1 + z_{\rm d})}{c}  \tilde{\psi}(\theta),
\end{eqnarray}
where $z_{\rm d}$ is the lens plane redshift and $\tilde{\psi}$ is defined as the deflection potential and its deflecting structure relies on $\theta$ \citep{Wagner2020} expressed as:
\begin{eqnarray}\label{deflection_potential_0}
\tilde{\psi}(\theta) = \frac{1}{(1 + z_{\rm d})}\frac{r_{\rm e}c^2}{2\pi}\frac{N_{\rm e}(\theta)}{ \nu^2},
\end{eqnarray}
where $N_{\rm e}(\theta)$ represents the projected electron density along the line of sight and $r_{\rm e}$ is the classical electron radius.

The total delay time is then
\begin{eqnarray}\label{total_delay}
\nonumber t_{\rm tot} &=& t_{\rm g} + t_{\psi} \\
&=& \frac{(1 + z_{\rm d})}{c}\frac{d_{\rm ol}d_{\rm os}}{d_{\rm ls}} \bigg[\frac{\alpha(\theta)^2}{2} + \psi(\theta) \bigg], 
\end{eqnarray}
where the dimensionless deflection potential $\psi(\theta)$ is defined as
\begin{eqnarray}\label{deflection_potential}
\psi(\theta) = \frac{d_{\rm ls}}{d_{\rm ol}d_{\rm os}} \tilde{\psi}(\theta),
\end{eqnarray}

\subsection{Lensing geometry and magnification}\label{subsec2.3} 

For a thin plasma lens, the geometric relationship between source and image positions follows an equation analogous to gravitational lensing \citep{Schneider1992,Bozza2001}, expressed as
\begin{eqnarray}\label{lens_equation}
\beta = \theta - \alpha,
\end{eqnarray}
where $\beta$ is the source angular position and $\alpha$ is the deflection angle. The sign of $\alpha$ determines whether rays converge ($\alpha > 0$), diverge ($\alpha < 0$), or remain undeflected ($\alpha = 0$).

Applying Fermat's principle ( $\nabla_{\theta} t_{\rm tot} = 0$), the deflection angle becomes
\begin{eqnarray}\label{deflection_angle}
\alpha = \nabla_{\theta} \psi(\theta) = \frac{1}{(1 + z_{\rm d})} \frac{d_{\rm ls}}{d_{\rm os}d_{\rm ol}}\frac{r_{\rm e}c^2}{2\pi}\nabla_{\theta}\bigg[\frac{N_{\rm e}(\theta)}{ \nu^2}\bigg].
\end{eqnarray}
Based on Equation (\ref{deflection_angle}), $\theta$ is generally taken as being no correlation with $\nu$,leading to $\nabla_{\theta} [N_{\rm e}(\theta)/\nu^2]\approx  \nabla_{\theta}N_{\rm e}(\theta)/\nu^2$ and $\alpha \propto \nu^{-2} \nabla_{\theta} N_{\rm e}(\theta)$. However, this assumption creates a fundamental contradiction: while assuming frequency-independent image positions, the theory predicts frequency-dependent deflection angles and DMs.

For an axially symmetric lens, the magnification ($\mu$) of the image is given by the ratio of image to source solid angles, calculated through the Jacobian determinant of the lens mapping $\mu^{-1} = \det{(A)}$ \citep{Schneider1985}. Due to plasma-lensing effects, the magnification only depends on the solid angle of image at different frequency. Thus, we have
\begin{eqnarray}\label{magnification0}
    \mu^{-1} = \frac{\beta}{\theta} \frac{{\rm d}\beta}{{\rm d}\theta},
\end{eqnarray}

\subsection{Observational constraints and parameter selection}\label{subsec2.4} 

Although the axially symmetric electron distribution with the exponential and power-law models in a plasma lens have been widely described \citep{Clegg1998,Cordes2017,Vedantham2017,Er2018,Er2020}, there are no empirical expressions for the detailed density structure of the plasma with the relation between $N_{\rm e}$ and image position ($\theta$). In addition, similar images can be predicted from the each model of two lensing forms and the added parameter for the addition of a finite core with angular core radius required to be considered in the plasma lens with the power-law model \citep{Er2018,Wang2022}. Thus, we chose only to review the electron density distribution following a Gaussian profile \citep{Vedantham2017,Er2018}, expressed as
\begin{eqnarray}\label{Gaussian}
N_{\rm e}(\theta) = N_0 \exp\left(-\frac{\theta^2}{2\sigma^2}\right),
\end{eqnarray}
where $N_0$ is the maximum electron column density within the lens and $\sigma$ is the effective angular structure of the lensing width. Based on the assumptions in Section \ref{subsec2.1}, we took $N_{\rm e}(\theta)$ to be approximated by the projected electron density along the line of sight, denoted $N_{\rm e}(\theta) \approx \rm DM(\theta)$ and $N_0=\rm DM_0$.
The DM variations of repeaters range from $1 $ to $\rm 10\,pc\,cm^{-3}$ \citep{Li2021,Xu2022,Mckinven2023} and many of them have higher DM values from their host galaxies than those of the intergalactic medium \citep{Niu2022}. Thus, the maximum electron column density in plasma-lensing plane can be adopted as $\rm DM_0 \approx 10\, pc\, cm^{-3}$. 

According to observations \citep{Fedorova2019,Simha2023}, some FRBs could pass through the intervening galaxies along their propagation path before reaching observers. This can complicate the evaluation of distances between the observer, lens, and source (i.e. $d_{\rm os}$, $d_{\rm ol}$, and $d_{\rm ls}$) if a plasma lens appears in the intervening galaxies. However, investigations have revealed that the intergalactic medium of many localized FRBs exhibits lower DM contributions as compared to their host galaxies \citep{Macquart2020,Niu2022,Wu2024}. A statistical analysis of large sample data from the initial CHIME/FRB catalog indicates that the plasma medium in the host galaxy primarily contributes to the FRB scattering timescale \citep{Chawla2022}. This result is further supported by the analysis of the located FRBs' Faraday rotation measures (RMs) \citep{Sherman2023}. Based on these observations, we place the plasma lens within the FRB host galaxy, adopting a lens-source separation of $d_{\rm ls} \approx 1 \rm~ kpc$. For typical cosmological distances to repeating FRBs, we set $d_{\rm ol} \approx d_{\rm os} \approx 500 \,\rm Mpc$.

The physical scale of the lens can be constrained by several observations. FRB 20121102A shows a persistent radio counterpart with a projected size of $\lesssim 0.7$ pc \citep{Marcote2017}, while FRB 20180916B is associated with a v-shaped star-forming region spanning approximately $1.5 \rm \, kpc$ \citep{Marcote2020}. In our Galaxy, plasma-lensing structures range from a few astronomical units (AU) to tens of AU scales \citep{Brisken2010,Graham2011,Vedantham2017,Kerr2018,Sprenger2022}. Assuming similar environments around FRB sources, we adopted a characteristic lens size of $d_{\rm ol}\sigma = 200 \rm AU$.

To calculate the redshift of the plasma lens, we employed the Planck 2018 $\Lambda$CDM parameters \citep{Aghanim2020}, as follows:
\begin{itemize}
    \item dark energy density: $\Omega_{\Lambda} = 0.685$;
    \item matter density: $\Omega_{\rm m} = 0.315$;
    \item Hubble constant: $H_0 = 100h \rm ~km~ s^{-1} ~ Mpc^{-1}$, with $h = 0.6736$.
\end{itemize}

\subsection{Frequency-dependent effects}\label{subsec2.5} 

\begin{figure*}
\centering
  \includegraphics[width=17cm]{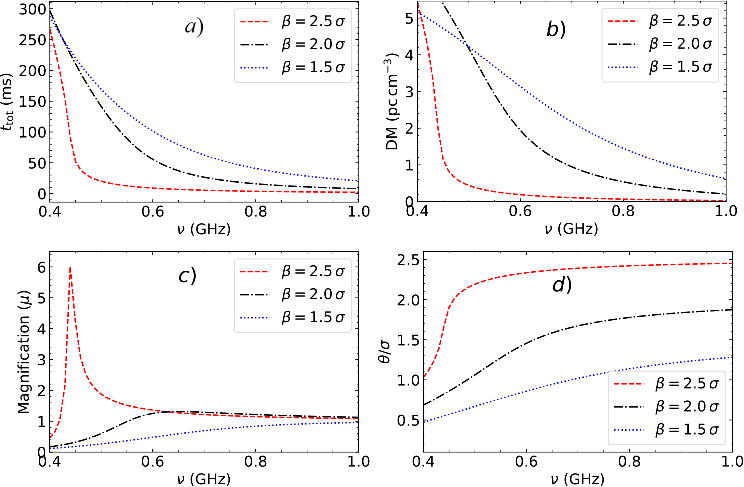}
  \caption{Frequency-dependent plasma-lensing effects for signals from three source positions ($\beta = 1.5\sigma$, $2.0\sigma$, and $2.5\sigma$; shown in blue, black, and red respectively). Panels show: a) Total delay time; b) Effective dispersion measure; c) Magnification; and d) Normalized image position ($\theta/\sigma$). }\label{fig2}
\end{figure*}

To investigate plasma-lensing effects across observationally relevant frequencies, we focus on the characteristic 0.4-1.0 GHz range, which is known for hosting a large number of CHIME/FRB's detection. Figure \ref{fig2} presents a comprehensive analysis of frequency-dependent lensing properties for three representative source positions ($\beta = 1.5\sigma$, $2.0\sigma$, and $2.5\sigma$). The results reveal systematic variations in four key parameters: total delay time, DM, magnification, and normalized image position ($\theta/\sigma$).

From this figure, we can find that signals at lower frequencies experience systematically lower magnification, while traversing regions of higher plasma density. This frequency-dependent behavior has profound implications for observed FRB properties. Not only do lower frequency signals undergo greater deflection, but they also experience the enhanced dispersion effects due to their interactions with denser plasma regions.

These findings challenge the traditional assumption of frequency-independent image positions ($\theta$). The conventional model posits that signals across all frequencies follow identical paths through the lens, encountering the same plasma distribution. However, our analysis demonstrates this simplification is incompatible with both the theoretical predictions (Figure \ref{fig1}) and observational evidence from pulsars and FRBs \citep{Cordes2016,Kaur2022,Wang2023}.

Figure \ref{fig1} contrasts these two possible lensing scenarios. In panel (a), our proposed framework demonstrates how signals at different frequencies ($\nu_1$ and $\nu_2$) naturally follow distinct paths through the lens, encountering varying electron densities along their trajectories. This behavior is not an intrinsic property of the plasma medium itself; rather, it emerges from the frequency-dependent paths taken by radio signals through spatial electron density gradients. Additionally, it could produce frequency-dependent DMs and arrival times that are consistent with observations. In contrast, panel (b) depicts the traditional model's frequency-independent path assumption, which fails to account for observed frequency-dependent effects.

The physical picture that emerges suggests a more complex interplay between frequency and plasma lens than previously recognized. As radio signals traverse the lens, their frequencies fundamentally influence their propagation paths, leading to the rich phenomenology observed in FRB signals. This frequency-dependent framework provides a natural explanation for various observed features, including frequency-dependent DMs and systematic arrival time variations. The detailed treatment of frequency-dependent effects in plasma-lensing models is discussed in the following section.

\section{Frequency-dependent dispersion measure}\label{sec3}
\subsection{Frequency-dependent DM and power-law DM model}\label{subsec3.1}

A detailed analysis of these signals reveals that DM values vary systematically with frequency, despite the use of consistent de-dispersion methods. This variation can be quantitatively characterized by fitting DM measurements across multiple frequency bands \citep{Cordes2016,Tuntsov2021,Kaur2022,Wang2023,Faber2024}. 
This phenomenon has been observed for both pulsars and FRBs. Observations of millisecond pulsar PSR J2241-5236 demonstrate clear power-law scaling between DM and frequency. Analyses using wide-band de-dispersion techniques yield $\rm DM\propto \nu^{2.54}$, while narrowband methods reveal $\rm DM\propto \nu^{3.02}$ \citep{Kaur2022}. This consistency across different analysis methods strengthens the evidence for intrinsic frequency dependence. Additionally, some repeating FRBs exhibiting the "sad trombone" effect show arrival times of sequential sub-pulses varying linearly with band center frequency \citep{Josephy2019}. These sequential sub-pulses likely represent components of a single burst that are differentially deflected by intervening plasma. Statistical analyses suggest the relationship between sub-pulse DM and center frequency follows $\rm DM(\nu)\propto \nu^2$ \citep{Tuntsov2021}.

Based on these observational constraints, we propose a general power-law model for frequency-dependent DM,
\begin{eqnarray}\label{equation5}
{\rm DM}(\nu) = A\nu^{\gamma}_{\rm GHz} + B,
\end{eqnarray}
where $\gamma$ is the frequency index, $\nu_{\rm GHz}$ represents frequency in GHz, $A$ controls the strength of the frequency dependence, and $B$ is for the maximum ($A<0$) or minimum ($A>0$) DM across the plasma-lensing plane. Here, $A$ and $B$ are constants.

\subsection{Modified deflection theory}\label{sec3.2}

\subsubsection{Deflection angle, $\alpha$}\label{subsec3.2.1}

Traditional plasma-lensing models assume a direct relationship between electron density and image position, $N(\theta)$ \citep{Er2018,Main2018}. However, theoretical analyses suggest that both image position and dispersion measure depend explicitly on frequency. We therefore propose a modified framework incorporating frequency-dependent effects through $\rm DM(\nu)$ and $\theta(\nu)$.

In this framework, the deflection angle with frequency-dependent effects can be expressed as 
\begin{eqnarray}\label{deflection_angle1} 
\alpha &=& \frac{1}{(1 + z_{\rm d})}\frac{d_{\rm ls}}{d_{\rm os}d_{\rm ol}}\frac{r_{\rm e}c^2}{2\pi\nu^2}\bigg[\frac{{\rm d} {\rm DM}(\nu)}{{\rm d} \nu} - \frac{2 {\rm DM}(\nu)}{\nu}\bigg]\frac{{\rm d}\nu}{{\rm d}\theta}. 
\end{eqnarray} 
This expression is updated form of Equation (\ref{deflection_angle}), and it implies how signal deflection arises from the interplay between the DM frequency scaling and mapping between frequency and image position. Several parameters in Equation (\ref{deflection_angle1}) can be constrained by observations ($d_{\rm os}$, $d_{\rm ol}$, $z_{\rm d}$) or estimated from source properties ($d_{\rm ls}$). The determination of ${\rm d}\nu/{\rm d}\theta$ requires additional physical constraints, introduced below.   

Unlike the classical approach where DM is assumed to be a function solely of position ($\rm DM(\theta)$), our framework explicitly incorporates the frequency dependence of the image position itself. The path followed by a signal of frequency, $\nu$, is determined by $\theta(\nu)$, resulting in an observed dispersion measure $\rm DM(\nu)$. This approach aligns with the physical picture shown in Figure \ref{fig1}a, where signals at different frequencies naturally follow distinct paths through the plasma lens.

Considering the plasma-lensing effect, the delay time between the pulses from a pulsar or the sub-pulses of a burst from a FRB at different center frequencies can be expressed as
\begin{eqnarray}\label{delay}
t_{\rm d} =\frac{cr_{\rm e}}{2\pi \nu_{0}^2} \frac{\rm DM(\nu)}{\nu^2_{\rm GHz}},
\end{eqnarray}
which could be caused by the geomitric effects of plasma lens \citep{Er2020}, i.e., $t_{\rm d} \simeq t_{\rm g}$. Coupled with Equations (\ref{geometric_delay}), (\ref{equation5}), and (\ref{delay}), the image position is given by
\begin{eqnarray}\label{equation6}
\theta(\nu) &=& \theta_0\frac{\left(A\nu_{\rm GHz}^{\gamma} +B\right)^{1/2}}{\rm DM_0^{\prime 1/2}\nu_{\rm GHz}} + C,
\end{eqnarray} 
where $\rm DM_0^\prime = 1 \, pc \, cm^{-3}$, and $C$ is a constant. Then, $\theta_0$ is defined as the characteristic angular scale and depends on the plasma-lensing parameters ($d_{\rm os}$, $d_{\rm ol}$, $d_{\rm ls}$, $z_{\rm d}$) expressed as
\begin{eqnarray}\label{D}
\theta_0 = \frac{1}{(1+z_{\rm d})^{1/2}} \left(\frac{d_{\rm ls}}{d_{\rm os} d_{\rm ol}} \right)^{1/2}\frac{r_{\rm e}^{1/2} c}{\nu_0(\pi)^{1/2}} \rm DM_0^{\prime1/2},
\end{eqnarray}
where $\nu_0 = 10^9 \, \rm GHz$. 

According to Equations (\ref{geometric_delay}) and (\ref{delay}), the deflection angle should be written as
\begin{eqnarray}\label{deflection_angle3}
\alpha = \theta_0\frac{\left(A\nu_{\rm GHz}^{\gamma} +B\right)^{1/2}}{\rm DM_0^{\prime 1/2}\nu_{\rm GHz}}.
\end{eqnarray}
Substituting Equations (\ref{equation6}) and (\ref{deflection_angle3}) into the lensing Equation (\ref{lens_equation}), we can derive the source position $\beta = C$. Consequently, it is evident that plasma lens with power-law DM could result in different magnification compared to classical plasma lens.

\subsubsection{Magnification derived from geometric graph}\label{subsec3.2.2}

\begin{figure}
\centering
  \resizebox{\hsize}{!}{\includegraphics{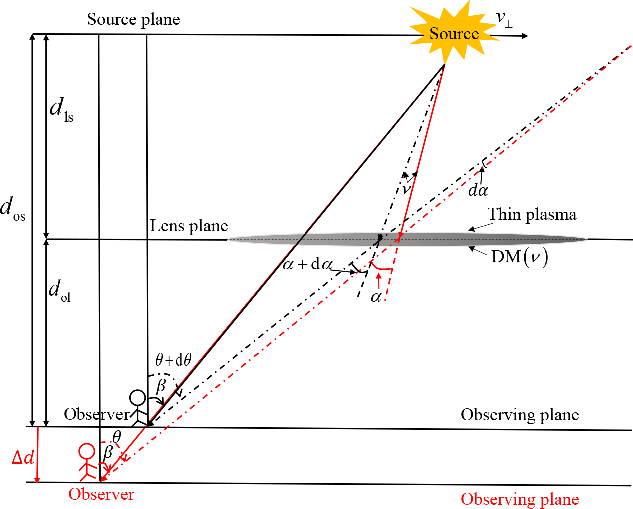}}
  \caption{Geometric diagram illustrating magnification calculation. Radio signals at a given frequency passing through adjacent image positions ($\theta$ to $\theta + d\theta$) map to different observer positions, with the magnification determined by the ratio of subtended solid angles.}\label{fig3}
\end{figure}

Considering radio signals at a fixed frequency pass through adjacent image positions from $\theta$ to $\theta + {\rm d}\theta$ (Figure \ref{fig3}). These signals map to an area element $d_{\rm ol}^2\theta {\rm d}\theta$ in the lensing plane. For closely spaced observers (separation $\Delta d$), the corresponding area in the observation plane becomes $d_{\rm ol}^2\beta {\rm d}\alpha$. This geometric mapping leads to a modified magnification of
\begin{eqnarray}\label{magnification1}
\nonumber \mu_{\rm new}^{-1} &\simeq& \frac{\beta}{\theta} \frac{\rm d\alpha}{\rm d \theta}\\
     &=& \left[ \frac{\theta_0}{C} \frac{(A\nu_{\rm GHz}^{\gamma} + B)^{1/2}}{\rm DM_0^{\prime 1/2}\nu_{\rm GHz}} + 1\right]^{-1}.
\end{eqnarray}

Comparing with the classical approaches, Equation (\ref{magnification1}) reduces the number of free parameters and is direct connection to observable quantities. Moreover, this formula explicit incorporation of frequency dependence and it gives natural explanation for frequency-dependent magnification effects. The remaining uncertainty in parameter $C$ represents the characteristic angular scale of the lensing core, potentially constrainable through detailed observations of individual FRB sources. This framework provides a more complete description of frequency-dependent plasma-lensing theory, while maintaining analytical tractability.

This revised magnification formula explicitly captures how frequency-dependent paths through plasma structures affect observed signal intensities. Unlike classical approaches, which struggle to self-consistently model frequency-dependent effects, our framework establishes a direct connection between the power-law scaling of frequency-dependent DM and the resulting magnification variations across frequency. This naturally explains why lower frequency components of FRB signals often appear systematically fainter as they traverse regions of higher electron density, which both disperse and de-magnify the signal through geometric effects.

\section{Numerical results and application to FRBs}\label{sec4}

The formation of a sequence of narrowband sub-pulses in a FRB could be the result of deflection effects caused by variations in the plasma medium on the lensing plane along the propagation path \citep{Tuntsov2021}. Specifically, the deflection angle exhibits frequency-dependent properties for sub-pulses, indicating a possible broadband emission mechanism for FRBs. On this basis and the assumptions in Section \ref{subsec2.1}, the "sad trombone" effect can be attributed to components of the same burst being deflected differently at various frequencies by spatially varying plasma media across the lensing plane. This result leads to existing differences in geometric trajectories and DMs between the sub-pulses at different frequencies, causing delays in their arrival times \citep{Wang2022}. 

However, this scenario suggests that the FRB could originate from a broadband emission mechanism, consistent with various models for FRB origins \citep{Zhang2023}. The narrowband property of sub-pulses could be caused by approximately uniform plasma clouds in the local lensing region, similar to but different from the model of \cite{Tuntsov2021}. Therefore, the DMs of all sub-pulses exhibit a frequency correlation overall, implying that the bursts of an FRB statistically have frequency-dependent DMs.

In this section, we only focus on the frequency-dependent magnification of FRBs due to the plasma lens with the frequency-dependent DM. This effect can explain the frequency-dependent luminosity of sub-pulses \citep{Gajjar2018,Amiri2019,Amiri2020,Luo2020}, the complex spectral morphologies of sub-pulse components observed in non-repeating FRBs \citep{Faber2024}, and other related phenomena. We applied this magnification to describe the properties of bursts for FRB 180814.J0422+73, but to discuss other effects caused by this type of plasma lens in future research. Moreover, the image position pattern and magnification of multiple plasma lenses have properties similar to that of the single lens \citep{Er2022b}, we only need to consider a single lens here. 

\subsection{Numerical results of plasma lens with $\rm DM \propto \nu^{\gamma}$}\label{subsec4.1}

For a plasma lens, we need to select suitable parameter values for our discussions. Firstly, $\theta_0$ can be set based on the parameters given in Section \ref{subsec2.4}. Secondly, $\gamma$ can take values of 2.0, 2.5, or 3.0, based on the observations of the millisecond pulsar PSR J2241-5236 \citep{Kaur2022} and repeaters \citep{Tuntsov2021}. The value $B \simeq 10 \, \rm pc \, cm^{-3}$ is derived from the DM variations of these repeaters, while $A \simeq -1.00$, $-0.10$, or $-0.01$ is chosen based on the lowest and highest center frequencies in the observations of PSR J2241-5236 \citep{Kaur2022} and the majority of FRBs discovered around 1 GHz \citep{Petroff2016,Amiri2021}. 

Observations of Crab demonstrate that plasma lens could occur due to the dense filaments in the Crab Nebula \citep{Graham2011}. The optical filaments of the Crab Nebula indicate larger sizes of plasma medium, of approximately $1000\,\rm AU \times 0.5\,\rm pc$ \citep{Davidson1985}, whereas the possible scale structure of plasma lens is from a few AU to tens of AU \citep{Brisken2010,Graham2011,Kerr2018,Sprenger2022} or possibly larger \citep{Vedantham2017}. Therefore, we took $C \simeq -10 \theta_0$, $-20 \theta_0$, or $-50 \theta_0$. 

\begin{figure*}
\centering
  \includegraphics[width=17cm]{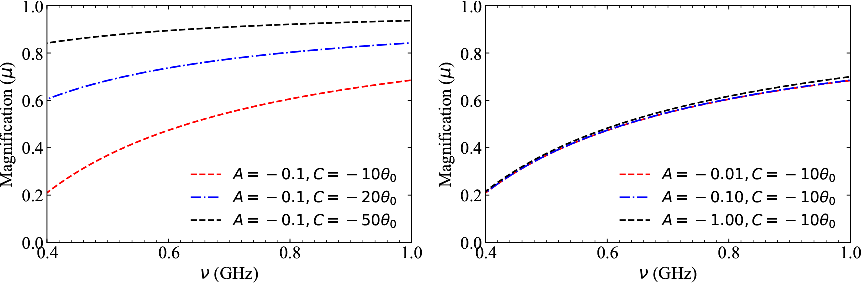}
  \caption{Frequency-dependent magnification for plasma lens with $\gamma = 2$. Left panel: Magnification curves for three source positions ($C = -50\theta_0$, $-20\theta_0$, and $-10\theta_0$) with $A = -0.1$, $B = 10\, \rm pc\, cm^{-3}$. Right panel: Magnification curves for a fixed source position ($\beta = -20\theta_0$) with different lens strengths ($A = -1.00$, $-0.10$, and $-0.01$).}\label{fig4}
\end{figure*}

\begin{figure}
\centering
  \resizebox{\hsize}{!}{\includegraphics{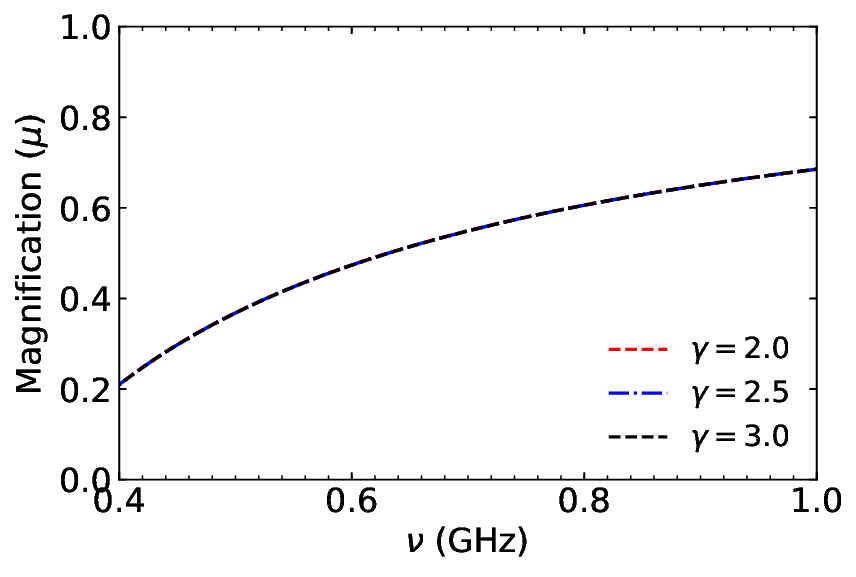}}
  \caption{Comparison of magnification curves for different frequency power-law indices ($\gamma = 2.0$, 2.5, and 3.0) in the plasma-lensing model. All curves calculated for source position $C = -10\theta_0$ with $A = -0.1$ and $B = 10 \,\rm pc\, cm^{-3}$.}\label{fig5}
\end{figure}

Based on assumptions in the Section \ref{subsec2.1}, the magnification of FRBs could approximate invariant in burst's duration with time. Moreover, the medium of plasma lens has quite weak radiation, implying the relation between image position and DM is difficult to obtain via observations, but the image position can be inferred by frequency-dependent DM and residual delay time in de-dispersed pulses. In this work, we only discuss the relation between the frequency and magnification, as showed in the Figs. \ref{fig4} and \ref{fig5}. Thus, one-dimensional (1D) images are enough to explore isotropic equivalent luminosity of some FRBs varied with frequency. For other properties of some FRBs undergone latent plasma-lensing effects, our scenario might have to be extended to two-dimensional (2D) images in the future.

As exihibited in the Fig. \ref{fig4}, the magnification decreases as the frequency decreases when the radio signals originate from the same source. Radio signals from greater $|\beta| = |C|$ have a larger magnification, as shown in the left panel of Fig. \ref{fig4}. However, radio signals from the same source position show the sight variations at high frequencies when affected by different plasma lenses (different values of $A$ in Equation (\ref{equation5})), as shown in the right panel of Fig. \ref{fig4}. We also give the magnification due to the plasma lens with varied frequency indices in Fig. \ref{fig5}, revealing a nearly same relation between magnification and frequency. These results suggest that the magnification of radio signals caused by a plasma lens with $\rm DM \propto \nu^{\gamma}$ is mainly dependent on their frequency and source position. 

\subsection{Plasma-lensing effects on the repeating FRB 180814.J0422+73}

In Section \ref{subsec4.1}, we find that the magnification of light rays at different frequencies could be frequency-dependent when they pass through the plasma lens for our scenario. In addition, for the bursts of some repeaters with sub-pulse components drifting downward with their center frequencies in a linear proportion, each sub-pulse component could also propagate along different paths and traverse through different plasma mediums, resulting in the relationship of $\rm DM \propto \nu^2$ \citep{Tuntsov2021}. Thus, the observed magnification could be primarily come from the combined effects of the plasma lens and the radiation mechanism of source.

Assumed this type of burst with $N$ sub-pulses is suffering from the plasma-lensing effects, we label the sub-pulses from the highest to lowest frequencies as the first to $N$th sub-pulses, respectively. The $n$th sub-pulse is denoted with subscript $n$ in its parameters. The ratio of isotropic luminosity between the $n$th and first sub-pulse is expressed as $L_{{\rm iso},n}/L_{\rm iso,1}$. To account for the plasma-lensing effects, the ratio of signals detected by observers ($R$) must be multiplied by the magnification ratio, $ \mu_{n}/\mu_1 $. Consequently, the final ratio is
\begin{eqnarray}\label{ratio}
R \simeq \bigg|\frac{\mu_{n}}{\mu_1}\bigg|\frac{L_{{\rm iso},n}}{L_{\rm iso,1}}.	
\end{eqnarray}

The fluence of the burst is defined as the average product of its peak flux density and width, closely matching the result given by \citealt{Petroff2016}. Therefore, the observed equivalent isotropic luminosity can be approximated as
\begin{eqnarray}\label{anisotropic_energy}
L_{{\rm iso,}n}^{\rm obs} \approx 8\pi P_n(\nu_{{\rm GHz,}n}) d_{\rm os}^2 \Delta \nu_{n},
\end{eqnarray}
where $P_{n}$, $\nu_{{\rm GHz,}n}$, and $\Delta \nu_{n}$ represent peak flux density, center frequency, and spectral bandwidth of the observation, respectively.

We use the bursts of FRB 180814.J0422+73 as an example to explore the effects of plasma lens. This repeater has had bursts at barycentric MJD 58378.03504878 and MJD 58419.43090504 with more than two sub-pulses \citep{Amiri2019}. The sub-pulses of each burst, de-dispersed by $\rm DM = 189.4\, \rm pc\, cm^{-3}$, showed a linear downward drift in their center frequencies, with drift rates of $-6.4\pm 0.7 \,\rm MHz\, ms^{-1}$ and $-1.3\pm 0.3 \,\rm MHz\, ms^{-1}$, respectively. This suggests that the FRB could follow the relation $\rm DM \propto \nu^2$, allowing us to calculate the corresponding $B = 6.62\,\rm pc \, cm^{-3}$ and $8.67\,\rm pc \, cm^{-3}$ based on their center frequencies. The parameters of the sub-pulses of FRB 180814.J0422+73, including the center frequencies, bandwidths, ratio of the peak flux density between the $n$th and first sub-pulses, and the ratio of isotropic luminosity given by Equation (\ref{anisotropic_energy}), are listed in Table \ref{FRB_180814}. 

According to theoretical predictions, the isotropic luminosity of the bursts at different frequencies is determined by the radiation mechanism of a repeater \citep{Kumar2017,Beloborodov2020,Zhang2022}. This implies that $L_{\rm iso} \propto \nu_{\rm GHz}^{\delta}$, where $\delta$ can take values of $2/3$, $0$, $-1$, $-4/3$, or $-6$. We can then combine the isotropic luminosity ratios of the sub-pulse bursts at barycentric MJD 58378.03504878 and MJD 58419.43090504 with Equations (\ref{magnification1}), (\ref{ratio}), and (\ref{anisotropic_energy}). The fitting results, including parameters $A$, $\theta_0/C$, and R-square values, are presented in Tables \ref{FRB_180814_2} and \ref{FRB_180814_3}.

\begin{table*}
  \centering
  \caption{Properties of FRB 180814.J0422+73 sub-pulses observed at barycentric MJD 58378.03504878 and MJD 58419.43090504 \citep{Amiri2019}.}\label{FRB_180814}
  \begin{center}
  \begin{tabular}{llccccc}
\hline\hline\noalign{\smallskip}
   Burst                               &  DM                                  &  Component & Center frequency       & Bandwidth  & $P_{n}/P_1$   & $L_{{\rm iso},n}^{\rm obs}/L_{\rm iso,1}^{\rm obs}$ \\ 
                                            &   ($\rm pc \, cm^{-3}$ )  &                          &           (GHz)                   &    (MHz)         &                            &                                                                                                        \\    
  \hline\noalign{\smallskip}
   MJD 58378.03504878 &   189.5                              &1                      & 0.775                              & 83                 & 1.000                &1.000                                                                                               \\     
                                            &                                           &2                      & 0.711                              & 85                 &0.760                 &0.778                                                                                               \\  
                                            &                                           &3                      & 0.636                              & 95                 &0.250                 &0.286                                                                                               \\   
   MJD 58419.43090504 &  188.9                              &1                      & 0.485                              & 55                 &1.000                 &1.000                                                                                               \\   
                                            &                                           &2                      & 0.468                              & 54                 &0.768                 &0.754                                                                                                \\   
                                            &                                           &3                      & 0.455                              & 58                 &0.928                 &0.979                                                                                                \\  
                                            &                                           &4                      & 0.433                              & 45                 &0.732                 &0.599                                                                                                \\   
                                            &                                           &5                      & 0.422                              & 34                 &0.393                 &0.243                                                                                                 \\   
  \noalign{\smallskip}\hline\hline
  \end{tabular}
  \tablefoot{The columns are the DM, center frequency, bandwidth, and relative peak flux density ($P_n/P_1$) and isotropic luminosity ($L_{{\rm iso},n}^{\rm obs}/L_{\rm iso,1}^{\rm obs}$) for each subpulse.
  }
  \end{center}
\end{table*}

The burst at barycentric MJD 58378.03504878 has three sub-pulses. We find that the fitting value of $\theta_0/C$ significantly affects the R-square, but there is an insignificant increase in R-square with increasing $A$ when $-1 < A < 0$. In addition, observations of the millisecond pulsar PSR J2241-5236 give $A \sim -10^{-4} \,\rm pc\, cm^{-3}$ \citep{Kaur2022}, and FRB 20180916B between the center frequencies $350\, \rm MHz$ and $600 \, \rm MHz$ has DM variations less than $0.3\, \rm pc\, cm^{-3}$ \citep{Wang2023}. Taking $A = -0.01$, we find that combining possible radiation scenarios with plasma-lensing effects yields an R-square greater than 0.8 when the fitting parameter $\theta_0/C \sim -0.235$.

For the burst at barycentric MJD 58419.43090504, the five sub-pulses are fitted with $\theta_0/C \sim -0.138$. The radiation mechanisms with $\delta = 2/3$ or $\delta = 0$ bring R-square values less than 0.7; whereas those with $\delta = -4/3$ or $\delta = -6$ yield R-square values greater than 0.7. The results above suggest a higher probability that the repeater is aligned with models where the isotropic luminosity index $\delta = -4/3$ or $-6$.

The lower R-square could be due to the second sub-pulse having an abnormal ratio of isotropic luminosity ($L_{{\rm iso},2}^{\rm obs}/L_{\rm iso,1}^{\rm obs}$). This sub-pulse exhibits a slightly wider pulse width than the others \citep{Amiri2019}, indicating possibly higher scattering and absorption effects due to the plasma medium it traversed. We infer that the relatively strong plasma absorption effects could be due to a distinct axial ratio deviating from 1 for the second sub-pulse, resulting in a lower ratio of isotropic luminosity compared to the other four sub-pulses \citep{Draine2011,Vedantham2017}.

When the second sub-pulse is excluded, the remaining sub-pulses, combined with plasma-lensing effects and possible isotropic luminosity models, can be fitted using the same $A$ and $\theta_0/C$ as those given in Table \ref{FRB_180814_3}. The results show that models with isotropic luminosity indices of $\delta = 2/3$ and $0$ still arise R-square values less than 0.7, but the R-square values increase to 0.723 and 0.974 for scenarios with isotropic luminosity indices of $\delta = -4/3$ and $-6$, respectively. The observed frequency-dependent luminosity scaling, when accounting for plasma-lensing effects, appears most consistent with emission models involving magnetospheres of magnetized compact objects \citep{Kumar2017,Zhang2023}. However, alternative explanations cannot be definitively ruled out without additional multiwavelength and polarization data. We await the results of future FRB studies with multiwavelength and polarization observations to future explore and identify this burst.

\begin{table}
  \centering
  \caption{Fitting results for FRB 180814.J0422+73 burst at MJD 58378.03504878.}\label{FRB_180814_2}
  \begin{center}
  \begin{tabular}{lccccc}
  \hline\hline\noalign{\smallskip}
     $L_{\rm iso}$                   &  $A$  & $\theta_0/C$     &   R-square   \\
  \hline\noalign{\smallskip}
   $\propto \nu^{2/3}$               &-0.010        &-0.230               & 0.956            \\   
   $\propto \nu^{0}$                   &-0.010       &-0.229               & 0.982            \\     
   $\propto \nu^{-4/3}$             &-0.010        &-0.235               & 0.999             \\   
   $\propto \nu^{-6}$                 &-0.010        &-0.245               & 0.847             \\   
  \noalign{\smallskip}\hline\hline
  \end{tabular}
  \tablefoot{Results show plasma lensing parameters ($A$, $\theta_0/C$) and goodness-of-fit (R-Square) for different intrinsic luminosity scaling relations
  }
  \end{center}
\end{table}

\begin{table}
  \centering
  \caption{Similar to Table \ref{FRB_180814_2}, but for the burst at MJD 58419.43090504.}\label{FRB_180814_3}	
  \begin{center}
  \begin{tabular}{lcccc}
  \hline\hline\noalign{\smallskip}
     $L_{\rm iso}$                   &  $A$  & $\theta_0/C$     & R-square \\
  \hline\noalign{\smallskip}
   $\propto \nu^{2/3}$               &-0.010         &-0.136              & 0.630    \\   
   $\propto \nu^{0}$                   &-0.010        &-0.137               & 0.658  \\     
   $\propto \nu^{-4/3}$              &-0.010        &-0.138              &  0.719    \\ 
   $\propto \nu^{-6}$              &-0.010        &-0.141              &  0.862    \\
  \noalign{\smallskip}\hline\hline
  \end{tabular}
  \tablefoot{Results demonstrate how different intrinsic luminosity models combined with plasma lensing effects explain the observed frequency-dependent flux variations.
  }
  \end{center}
\end{table}

\section{Conclusion and discussion}\label{sec5}

The classical Gaussian plasma-lensing model demonstrates that the properties of radio signals, including delay time, DM, magnification, and image position, exhibit a strong frequency dependence when signals at different frequencies originate from the same source \citep{Wang2022}. Our analysis reveals that radio signals at different frequencies follow distinct propagation paths, characterized by $\theta(\nu)$ and $\rm DM(\nu)$ rather than the traditional frequency-independent $\rm DM(\theta)$. In addition, the magnification of the image varies with the frequency of the radio signals. 

Observations show that frequency-dependent DMs follow a power-law relationship between the image's angular position and frequency \citep{Kaur2022,Faber2024}. By using this relationship, we derived equations for frequency-dependent the image position and deflection angle, demonstrating that light rays of different frequencies traverse different propagation paths, while the source position remains constant. From our geometric analysis, a new magnification equation emerges to show that plasma-lensing magnification decreases with decreasing radio signal frequency. The frequency-dependent DM model presented in this work offers a physical mechanism for understanding the frequency-dependent properties of some FRB signals, which are difficult to explain within standard dispersion frameworks. By explicitly incorporating frequency-dependent paths through structured plasma environments, our model successfully predicts systematic variations in signal's magnification that are well aligned with observations, as demonstrated by our analysis of FRB 180814.J0422+73. Future multifrequency observations should reveal clear signatures of this effect, including systematic variations in measured DM values across frequency bands, as well as consistent correlations between DM, magnification, and arrival times that follow predicted power-law relationships.

The behavior of FRB 180814.J0422+73 appears to be fittingly explained by our model. Specifically, the peak flux densities of sub-pulses for this repeater are decreasing with the decreasing center frequency when combined effects between its intrinsic properties and the effects of plasma lens are taken into account. In addition, this requires for the bursts of the repeater to be produced from the magnetized compact star's magnetosphere. Similar plasma-lensing signatures appear in other FRBs, including FRB 20121102A's burst at MJD 57991.409904044 \citep{Levkov2022} and several other sources showing characteristics comparable to FRB 180814.J0422+73 \citep{Amiri2019,Amiri2020,Luo2020,Levkov2022}. 

These findings have significant implications for FRB physics and observation. A perturbed plasma along the line of sight causes FRBs to deviate from classical dispersion relations \citep{Er2020,Wang2022} and produces frequency-dependent magnification effects \citep{Er2018,Wang2022}. This behavior provides new tools for constraining plasma-lensing models, electron density distributions, and intrinsic source properties. In particular, a number of repeating FRBs, including FRB 20121102A, FRB 20180916B, and FRB 20201124A, show evidence of plasma clump obscuration \citep{Chatterjee2017,Marcote2020,Xu2022} with associated radio emission \citep{Tendulkar2017,Li2022,Ravi2022}.

The frequency-dependent propagation paths implied by plasma lens suggest that the DMs for repeating FRBs vary with the observed frequency. Some repeaters show evidence of such frequency-dependent paths \citep{Cho2020,Feng2022,Wang2023}, while the delay time still resides in their signals between high and low frequencies for some non-repeaters after de-dispersion by taking classical disperstion relation is done \citep{Faber2024}. This indicates that some repeating and non-reapting FRBs could propagate along possible frequency-dependent paths. However, direct observations of frequency-dependent DMs for these sources remain rare. If such DM variations are detected by multifrequency observations in the future, it would provide crucial insights into FRB radiation mechanisms after accounting for plasma-lensing effects \citep{Kumar2017, Beloborodov2020}.

The high event rate of FRBs (818 $\rm day^{-1}\,sky^{-1}$ at 600 MHz) reported by CHIME \citep{Amiri2021} provides an excellent opportunity for systematic studies of plasma-lensing effects. High-resolution multifrequency monitoring campaigns of repeating FRBs will be important for tracking DM variations and mapping plasma medium structures in detail. These observations, combined with improved statistical analysis for parameter estimation and model discrimination, will enable more constraints on both lensing properties and source characteristics.

Our theoretical framework can be extended in several important ways. Future models should incorporate non-axisymmetric geometries and time-varying effects, which might be more effective in presenting the complex plasma environments around FRB sources. The frequency-dependent magnification effects observed in many FRBs suggest that plasma lens could serve as a valuable probe of both source environments and host galaxy properties. Indeed, the precise dispersion characteristics of FRBs are increasingly recognized as powerful tools for studying plasma properties across various scales, from compact object magnetosphere to galactic halos and HII regions \citep{Yang2017,Prochaska2019, Tsupko2020}.

As observational capabilities continue to improve, particularly with next-generation facilities such as SKA and enhanced CHIME, our framework will become increasingly valuable for extracting physical information from FRB observations and constraining source models. This approach promises to provide new insights into the nature of FRB sources and the properties of their host environments, advancing our understanding of these enigmatic phenomena.

\begin{acknowledgements} 

This work is supported by the National Key R\&D Program of China No. 2022YFA1603104, the Natural Science Foundation of Xinjiang Uygur Autonomous Region(No. 2023D01E20), the National Natural Science Foundation of China (No. 12403054, 12288102, 12033001, 12273028), Sichuan Provincial Natural Science Foundation Project (No.2025ZNSFSC0878), the talent introduction program of Sichuan University of Science \& Engineering (No.2024RC15), the Tianshan talents program (2023TSYCTD0013), the Major Science and Technology Program of Xinjiang Uygur Autonomous Region (No. 2022A03013-1, 2022A03013-3, 2022A03013-4), the Natural Science Foundation of Xinjiang Uygur Autonomous Region (No. 2022D01A363), and the Urumqi Nanshan Astronomy and Deep Space Exploration Observation and Research Station of Xinjiang (XJYWZ2303). AK acknowledges the support of the Tianchi Talents Project of Xinjiang Uygur Autonomous Region. We also acknowledge the support of science and technology department of gansu province (No. 20JR5RA481).
\end{acknowledgements}

\section*{Data availability}

Observational data used in this paper are quoted from cited works. Additional data generated from the computations can be made available upon reasonable request.


\end{document}